\documentstyle[12pt,aaspp4]{article}
\lefthead{Dolphin}
\righthead{HSTphot}
\begin{document}

\title{WFPC2 Stellar Photometry with HSTphot}

\author{Andrew E. Dolphin}
\affil{National Optical Astronomy Observatories, P.O. Box 26372, Tucson, AZ 85726\\
Electronic mail: dolphin@noao.edu}

\begin{abstract}
HSTphot, a photometry package designed to handle the undersampled
PSFs found in WFPC2 images, is introduced and described, as well as some
of the considerations that have to be made in order to obtain accurate
PSF-fitting stellar photometry with WFPC2 data.  Tests of HSTphot's 
internal reliability are made using multiple observations of the same
field, and tests of external reliability are made by comparing with
DoPHOT reductions of the same data.
\end{abstract}


\section{Introduction}

With the installation of WFPC2 in December 1993, Hubble Space Telescope
(HST) gained an imager capable of high-resolution stellar photometry.
This advance provided a number of opportunities, most notably the ability
to obtain deep photometry in crowded fields such as nearby galaxies and
globular clusters.  However, the severely undersampled WFPC2 point spread
function (the FWHM is comparable to a PC pixel and about half a WFC pixel)
produces a challenge when attempting to obtain accurate stellar photometry.
As a result, issues of the number of PSF points calculated per pixel by the
photometry software, and of approximations of quantum efficiency variations
and charge diffusion, which cause insignificant errors of well under a
percent in well-sampled data, contribute significant errors of order a
percent in PC data and greater in WFC data.

Because of the vast amount of WFPC2 data available to the astronomical
community, HSTphot was developed specifically for its reduction, allowing
for the creation of a highly specialized (and efficient) photometry
program.  The package (\textit{hstphot} and accompanying utilities) runs
from the Unix command line, and has been successfully compiled and run
on machines running Solaris and Linux.  As a manual is available
with the package, this paper is intended to describe and test HSTphot
rather than to give a detailed explanation of installation and use of
HSTphot.  Information on obtaining HSTphot can be obtained from the author
by e-mail or from the web.  As HSTphot is a continuing project,
contributions of improvements, additional utilities, and bug reports and
fixes are welcome.

Section 2 gives a description of the HSTphot algorithms, and points out
many of the differences between HSTphot and the well-known DAOPHOT and
DoPHOT packages.  This paper is intended more as a description of the
routines and choices that were made for this package and some of its
differences with DAOPHOT and DoPHOT, rather than as a detailed
description of techniques of stellar photometry.  The reader is encouraged
to read Stetson's (1987) introduction of DAOPHOT, which gives a very
thorough treatment of this subject.  Stetson et al. (1990) and Stetson
(1994) present later modifications to DAOPHOT, many of which are also
implemented in HSTphot.  It should also be noted that few, if any, of the
techniques and algorithms used by HSTphot are revolutionary.  Rather, I
am attempting to provide a photometry package in which the set of choices
made is as close to optimal as possible for WFPC2 stellar photometry.

Section 3 of this paper presents a series of tests that were run on
HSTphot photometry, examining its photometric and astrometric accuracy.
A comparison with a DoPHOT reduction of the same field is also given.

\section{HSTphot Algorithms}

As a PSF-fitting stellar photometry package, HSTphot is not significantly
different in concept from the well-known DAOPHOT (Stetson 1987) and
DoPHOT (Schechter, Mateo, \& Saha 1993) packages.  In order to obtain
calibrated photometry for an image, the following steps need to be
accomplished:
\begin{itemize}
\item Image Preparation
\item PSF Determination 
\item Detection of Stars
\item Iterative Photometry Solution
\item Aperture Corrections
\end{itemize}
It is assumed below that the data have had bias, dark current, and
flat-field corrections made before beginning this procedure, presumably
by the STScI pipeline.  The necessary files from the STScI archive for
HSTphot are the calibrated data image (c0f) and the data quality image
(c1f).  For clarification, utility names are given in italics in the
following sections.  For example, ``HSTphot'' refers to the entire
photometry package, while ``\textit{hstphot}'' refers to the specific
program that runs the photometry solution.

\subsection{Image Preparation}

A collection of image preparation utilities is provided with the HSTphot
package, which run the necessary processing steps such as masking of bad
columns, cosmic ray cleaning, hot pixel masking, etc.

The first preparation procedure is the masking of bad columns and pixels,
which is done with the \textit{mask} routine.  This simple routine will
read the data image (c0f) and the data quality image (c1f) provided by
STScI, and proceeds to mask out all pixels that are deemed to be bad -
types 1 (Reed-Solomon decoding error), 2 (calibration file defect), 4
(permanent camera defect), 16 (missing data), 32 (other bad pixel), 256
(questionable pixel), and 512 (unrepaired warm pixel).  This masking will
also eliminate the vignetted region in recent images (very early data
quality images do not flag this region).  Row 800 and column 800 are also
masked out entirely.  Because the saturation flag (type 8) in the data
quality image is unreliable, all pixels with 3500 or more counts are set
as saturated (4095 DN) to avoid ambiguity later in the reductions.  All
masked pixels are set to the bad data value, -100 DN, and are ignored for
the remainder of the photometry.

\subsection{Image Cleaning and Combination}

The masked image is then ready for cosmic ray cleaning and combination,
a process for which the utility \textit{crclean} is designed.  This
utility uses a routine based on the IRAF task CRREJ, itself a more
sophisticated version of the elementary ``maximum value reject'' method of
combining images.  \textit{Crclean} is provided a set of images to be
combined, which must be taken at the same pointing and with the same
filter, and compares the images at each pixel position.  All unmasked and
unsaturated pixels at that position are scaled for their respective
exposure times and compared.  The simple procedure would be to use the
either median or minimum value of the pixels at a given position as a
comparison value (\textit{crclean} provides both options for the user),
and reject all values that fall more than
\begin{equation}
\hbox{max deviation}=\frac{\sigma_{threshold}\sqrt{\hbox{Read Noise}^2+\hbox{counts}/\hbox{Gain}}}{\hbox{Exposure Time}}
\end{equation}
away from the comparison value.  In this equation, Read Noise and counts
are both expressed in DN for simplicity, and Gain is expressed in the usual
$e^{-}$ per DN.  The value $\sigma_{threshold}$ is user-defined, and is
the maximum number of standard deviations away for which a value should
be retained.  The recommended value of this parameter is near 3, in other
words a 3$\sigma$ threshold.  The pixel values meeting this criterion are
then added and scaled to produce an image with an effective exposure time
equal to the sum of the individual exposures.

The criterion above makes the assumption that all differences between the
images result from shot noise and cosmic rays.  In reality, of course,
slight image shifts, focus changes, and variability in the objects themselves
cause this assumption to be incorrect, and thus a more complex solution is
required.  All of the additional factors listed will produce a variability
between images that scales proportionally with counts rather than with the
square root of the counts, and thus such a term is added in quadrature to
the threshold,
\begin{equation}
\hbox{max deviation}=\frac{\sqrt{\sigma_{threshold}^2\hbox{Read Noise}^2+\sigma_{threshold}^2\hbox{counts}/\hbox{Gain}+c^2\hbox{counts}^2}}{\hbox{Exposure Time}}.
\end{equation}
The constant $c$ is user-defined, with values near 1 providing ``safe''
cosmic ray cleaning, causing values to be rejected only if they are at
least twice the comparison value (this is roughly what is required to
avoid chopping the peaks of stars in the case where star is centered on a
pixel in one image and centered at the corner of four pixels in another).
This extension of the threshold formula is also adapted from CRREJ.

Finally, in the case of poorly-aligned images (by a few tenths of a WFC
pixel), there will be a number of stars for which a pixel in one of the
images will contain significantly more of the total light than the same
pixel in the other images.  This happens on the side of the star, where the
PSF slope is the steepest.  To prevent this from damaging the stellar
images, a final additional level of complexity is added, following Saha et
al. (1996).  The lower allowable limit (comparison value minus the
threshold) from the previous paragraph is retained, but the upper limit is
modified as follows.  Rather than computing a comparison value and
threshold from the values of that pixel in all images, these values are
instead computed from the maximum values within a 3$\times$3 pixel square
centered on that pixel in all images.

A second optional cleaning step can be inserted in the reduction process
after the cosmic ray cleaning and background determination (below).  This
step will attempt to locate and mask hot pixels that were neither flagged
by the data quality image nor corrected by the STScI calibration pipeline.
Unfortunately, the value of such hot pixels is proportional to exposure
time and the count rate is stable between images, so \textit{crclean},
which simply compares the count rate in each pixel with that in other
images, will not detect them.
This utility, \textit{hotpixels}, is a very simple utility that masks all
pixels meeting both of the following criteria.  First, the sky-subtracted
pixel value is more than ten times the average of adjacent sky-subtracted
pixel values.  Second, the sky-subtracted pixel value is more than seven
standard deviations above the average value of adjacent pixels.  The intent
of this utility is not to locate and remove every hot pixel, rather only
those that are sitting on blank sky and thus very easy to detect.

As likely clear by the descriptions, both cleaning stages are
intentionally cautious in the pixels that are thrown away.  Given the very
sharp PSFs in the WFC images, this approach seems wise, as a star damaged
in the cleaning process is extremely difficult to fix, while most false
detections that escape the cleaning process will be identifiable in the
photometry output due to unusual $\chi$ or sharpness values.

\subsection{Background Determination}

The sky or background determination is the final mandatory pre-photometry
step of HSTphot, and is done with the \textit{getsky} utility.  Obviously
this is a task that could also be accomplished within \textit{hstphot},
but given the occasional need to re-run \textit{hstphot} with different
detection parameters it is preferable to have the sky determined only
once.  The sky value is calculated at each pixel, using the robust mean of
pixel values inside a square ``annulus'' centered on that pixel.  The
adoption of the square annulus, rather than the more typical round shape,
was made for computational ease (the less multiplication, the faster the
program runs).  Given that the inner ``radius'' is sufficiently far from
the pixel in question ($\sim$8 FWHM) that the shape is of little
consequence.  For the PC, the inner square is 33 pixels on a side and the
outer square 45 pixels, thus giving a maximum of 1064 pixels used for the
sky determination.  The WFC sky calculations use squares of half this size
(a 23 pixel outer square and a 19 pixel inner square), with a maximum of
240 pixels used.  The robust mean of all unmasked and unsaturated pixels
within this area, using a recursive rejection of pixels more than
$2.5\sigma$ below and $1.75\sigma$ above the mean value, is computed.
Convergence is determined when a pass rejects no pixels, and the mean sky
value from that final iteration is set as the sky value for the pixel.  In
order to ensure a smoothly-varying background, the sky image is boxcar
smoothed to determine the sky values that will be used by \textit{hstphot}.

A few comments regarding the sky calculation process are in order.  The
sky value can be calculated in one of three ways: a single calculation
before the photometry process, a calculation immediately preceding the
photometric measurement of a star, and a calculation simultaneous with
the photometric measurement.  The final choice is the most appealing, as
the $\chi^2$ fitting procedure should have little trouble in
distinguishing the flat sky from the variable stellar PSF and therefore
one can determine the ``true'' sky value underneath each star with ease.
However, as demonstrated by Stetson (1987) and confirmed in my own similar
experiments, the quality of photometry, as measured by the tightness of
CMD features, is degraded by this process because of the creation of an
additional free parameter.  Thus with the third choice eliminated, the
HSTphot package gives the option of using the first only or a combination
of the first and second.  The \textit{getsky} routine determines the
\textit{a priori} sky value at each pixel, while \textit{hstphot} can
determine a modified sky value for a star immediately before the photometry
solution and very close to the star, providing the user two good
alternatives for use depending on the condition of the data.

The selection of the sky region involves a tradeoff between three
characteristic size scales.  The inner radius must be large enough that
the star contributes a very small amount of light.  The inner ``radius'' of
9 WFC pixels was chosen with this in mind, although there is still a
measurable contribution of light ($\ge$0.007\% of the starlight per pixel
in the sky region, with the exact amount dependent on the filter and
temperature) from the star at this distance.  The outer radius must be
small enough that the background is either constant or can be fit by a
plane over the region used.  As most of my own projects involve photometry
of field populations in Local Group dwarf galaxies, which have essentially
a constant background, I chose to not be concerned with this restriction.
However, for the sake of consistency, the PC sky region was set to be
exactly twice the size of the WFC annulus, thus creating an identical
error in the case of rapidly-varying backgrounds.  Finally, the number of
pixels within the background region must be much larger than the size of
the region used for the photometry solution, so as to minimize the noise
caused by the finite sky area.  With a maximum photometry area of 81 WFC
pixels provided by the PSF library, the 240 pixel sky area ensures that any
scatter from the sky calculation will be no more than half the shot noise
from the sky inside the photometry region.

Properties of the optional $\delta$sky value that can be determined by
\textit{hstphot} (generally most useful for images with a rapidly-varying
background) are given below.  Use of this technique will tend to mimic
DoPHOT sky determinations, which also calculate a sky value extremely
close to the star.

\subsection{Point Spread Functions}

As with DAOPHOT and DoPHOT, \textit{hstphot}'s star brightness and position
measurements are made by determining the combination of those parameters
that best matches the observed data.  Consequently, \textit{hstphot} needs to
have a good estimate of the properties of a stellar image before it can
begin the photometry.  This is aided by the fact that, small focus changes
aside, the stellar PSF is consistent in all WFPC2 data taken with the same
filter, making it possible to build a library of typical WFPC2 PSFs to be
used as initial guesses when beginning photometry on an image.

Because of the undersampled PSFs, it was decided to calculate the synthetic
PSFs for a variety of subpixel centerings, with a spacing of every 0.2 PC
pixels (25 total centerings) and 0.1 WFC pixels (100 total centerings).
These PSFs were calculated using Tiny Tim PSFs, which were generated with
subsampling settings of 0.1 PC pixels and 0.05 WFC pixels for additional
resolution.  Next, a charge diffusion correction was applied to the Tiny
Tim PSF, which was equivalent to smoothing the subsampled PSF with a
Gaussian kernel of $\sigma=0.32$ pixels.  (The value of 0.32 pixels was
determined by trial-and-error, with this value producing the lowest
\textit{hstphot} median $\chi$ value for high signal-to-noise data.  The
value of 0.32 is probably accurate to within 0.05 pixels.)  Finally, a
subpixel quantum efficiency variation of roughly a 10\% efficiency
decrease from center to corner was applied, with this value determined in
the same way.  The choice of the QE fluctuation amount turned out to have
very little impact on the quality of the fits, and was thus difficult to
determine accurately, with any value between 5\% and 15\% returning
indistinguishable median $\chi$ values.

This process was repeated at 64 positions per chip (every 100$\times$100
pixels), thus generating a total of 1600 PSFs on the PC and 6400 PSFs
per WFC.  Again, it should be noted that the use of Tiny Tim PSFs is
not new to HSTphot; CCDCAP (Mighell \& Rich 1995), for example, uses Tiny
Tim PSFs to determine aperture corrections for small-aperture photometry.
However, HSTphot provides what is, to my knowledge, the most elaborate
application.

With the grid of PSFs by chip position and subpixel centering calculated,
it is worth examining the effect of using these quantized grids rather
than an analytic function (such as what is used by DAOPHOT and DoPHOT).
It should first be noted that the errors discussed here are of PSF shape
rather than the total PSF size.  Thus, while a PSF whose total number of
counts was in error by 1\% would create photometry with an error of 0.01
magnitudes, a PSF whose central pixel was in error by 1\% but whose total
size was correct would create a much smaller photometric error.  The
number of counts in a simple $\chi^2$ PSF fit minimization is
\begin{equation}
\hbox{counts}=\frac{\sum Residual \times PSF/\sigma^2}{\sum PSF^2/\sigma^2},
\end{equation}
which can be simplified to
\begin{equation}
\hbox{counts}=\frac{\sum Residual}{\sum PSF}
\end{equation}
and
\begin{equation}
\hbox{counts}=\frac{\sum Residual \times PSF}{\sum PSF^2}
\end{equation}
for limiting cases of extremely bright stars (in which the noise is
dominated by the photon noise of the star) and extremely faint stars (in
which the noise is dominated by the constant background and readout noise),
respectively.  Since the total number of PSF counts is correct, the
limiting-case bright star, whose photometric measurement is essentially
aperture photometry, is completely unaffected by PSF shape errors.  For a
limiting-case faint star with a typical WFC PSF with about 35\% of the
starlight in the central pixel, a PSF shape error in which the central
pixel of the model PSF is 1\% too bright will create an error of +0.007
magnitudes in the photometry.  Because of the $PSF^2$ term, a similar 1\%
error involving all four pixels adjacent to the center (which will
typically each include 9\% of the starlight) would create a magnitude
error of just +0.002 magnitudes.

Thus the magnitude errors created by the use of a quantized library of
PSFs can be determined based on the errors in the central pixel.  The WFC
chips, which have the sharpest PSFs and thus are the most likely to show
errors, are chosen for this analysis.  First, the effect of the
100$\times$100 pixel grid of PSFs is considered.  The typical ratio of the
central pixel values for diagonally adjacent chip positions is at most
1.016, implying a worst-case 0.8\% central pixel PSF error given the lack
of any interpolation in the chip positions.

Second, the error from the linear interpolation of PSFs to provide an
arbitrary subpixel centering can be estimated.  If one compares any PSF
in the library to the average of four PSFs diagonal from it, one finds
a typical error of 2.8\% in the calculated central pixel value.  Since
linear interpolation errors scale as the baseline squared, this implies
a typical central pixel PSF error of 0.7\% in the worst-case situation
in which the desired position is equally distant from all four nearby
library positions.

Finally, there is a very small error when interpolating near the center,
created by the fact that Tiny Tim calculates its PSFs centered on the
central pixel rather than on the corner of four pixels.  As the subsampling
factors are 10 on the PC and 20 on the WFC and I therefore chose to
combine blocks of 10$\times$10 and 20$\times$20 Tiny Tim pixels to create
the library PSFs, the true center of the best-centered PSFs is actually
0.07 pixels off-center in the PC and 0.035 pixels off-center in the WFCs.
The effect of this error can be determined by comparing a
perfectly-centered PSF to one interpolated from the library, with the
error in the central pixel value of the interpolated PSF determined to
be 0.3\%.

Thus three sources of error in the PSF library can be characterized,
and are combined for a worst-case error of 1.8\% in the central pixel
of the PSF, which will generate an error of 0.013 magnitudes in the
photometry of a limiting-case faint star.  However, the 1$\sigma$ scatter
in the photometry of the limiting-case faint star will be much smaller,
0.004 magnitudes.  Finally, it should be pointed out that our hypothetical
limiting-case star actually does not exist, as the limit of constant noise
is only reached if the star has zero counts.  Any actual detectable star
will contribute some amount to the noise, and thus have a smaller random
scatter than the 0.004 magnitudes calculated here.  Finally, it is worth
reiterating that the brightest stars are unaffected by these considerations.

Within \textit{hstphot}, the PSFs are modified further to compensate for
the geometric errors of geometric distortion and the 34th row error.  Both
of these factors decrease the effective pixel sizes, and thus it is
necessary to magnify the PSFs accordingly.  The geometric distortion
pixel sizes are calculated via the Holtzman et al. (1995a) distortion
correction equations; the 34th row error (noted by Shaklan, Sharman, \&
Pravdo 1995) is characterized by row heights calculated from data supplied
by Ron Gilliland.  The PSF magnification process is quite simple, with the
effective pixel width or height used to determine the fraction of the light
from a given row or column that should be moved into the adjacent row or
column.  To compensate for the fact that the actual PSF value on the outer
edge of a pixel is much less than the average PSF value within a pixel, and
thus the amount of light transferred should be less, the PC and WFC transfer
amounts are multiplied by 0.70 and 0.65, respectively.  This value is
determined for the typical central pixel, with a worst-case error of a
0.3\% in the corrected value of the central pixel, or a maximum magnitude
error of 0.002 magnitudes in our hypothetical zero count star.  Again, the
detailed correction of pixels other than the central pixel are of less
interest, given that the PSF$^2$ term in the photometric solution will
make such errors negligible.

It should be noted that the expansion of PSFs for these geometric errors
is a necessity for obtaining accurate PSF-fitting photometry, in addition
the the normally-recommended step of multiplying the image by the
effective pixel area map (cf Holtzman et al. 1995a).  While field-varying
PSFs from DAOPHOT or DoPHOT should compensate for the smoothly-varying
geometric distortion, the 34th row error needs to be compensated
similarly, something which, to my knowledge, is not done in any other
photometry package.  The use of an uncorrected PSF on an uncorrected image
will produce photometry for a star on an affected row in which the shape
matches well (thus producing negligible error for faint stars) but the
total number of counts detected is 2\% too large for bright stars, giving
a magnitude error of -0.02 magnitudes.  If the data are multiplied by the
row size image, the total number of counts are correct (giving correct
photometry of bright stars) but the central pixel has an error of 3\%,
producing a magnitude error of -0.02 magnitudes for the limiting-case
faint star.  Thus, while the latter case is preferable since the larger
random errors should minimize the effect of the systematic +0.02 magnitude
shift, in either case one will see a relative error of a few percent
between bright and faint stars.  Naturally, such an effect only matters
significantly for roughly 3\% of the stars, but it is my intention that
known (and correctable) systematic errors at more than the 1\% level be
eliminated.

\subsection{Star Detection}

The star detection algorithm is very similar to that used by DoPHOT
(Schechter et al. 1993).  After subtracting the sky image from the data
image, the residual is scanned for any peaks.  Like DoPHOT, a series of
passes are made through the image, first for the brightest peaks, then
slightly dimmer ones, etc.  At the location of any detected peak, an
initial guess is made for the central position of the star using the pixel
value-weighted averages of the positions of the peak and adjacent pixels,
\begin{equation}
X_{center}=\frac{\sum x_i (R_i-R_{min})}{\sum (R_i-R_{min})}\ \hbox{and}
\end{equation}
\begin{equation}
Y_{center}=\frac{\sum y_i (R_i-R_{min})}{\sum (R_i-R_{min})},
\end{equation}
where $x_i$ and $y_i$ are the X and Y values of the pixels used, $R_i$ is
the residual at each pixel, and $R_{min}$ is the minimum residual of any of
the pixels used in the centroid calculation.

Using the centroid position as an initial guess, \textit{hstphot} then runs
a photometry solution to improve the position and determine the star's
brightness.  Rather than running a formal nonlinear least-squares
procedure, the photometry solution is determined through an iterative
process.  At each trial position, a quality-of-fit parameter is determined
at the trial position and eight adjacent positions, with a stepsize of 0.2
pixels in the PC and 0.1 pixels in the WFC used to match the spacing of the
PSF grid.  If the current trial position provides the best fit, the
solution is considered converged; otherwise the best fit becomes the new
trial position and the process is repeated.

After convergence, if the signal-to-noise (defined in the next section) of
the star equals or exceeds the user-defined detection threshold, the star
is kept; otherwise it is rejected.  As noted above, this is nearly identical
to DoPHOT's star detection procedure, but is significantly different from
DAOPHOT's.  DoPHOT and HSTphot both scan the image for stars, and run a
photometry solution to determine if the star has a signal-to-noise at or
above the threshold value.  In contrast, DAOPHOT uses a convolution of the
image with a Gaussian of user-defined FWHM to detect locations where stars
appear to be.  Such a routine will fail in cases in which the PSF FWHM
changes significantly over the image and cases in which the background is
variable.  In either case, the DAOPHOT detection limits will be
poorly-defined, with a field-varying PSF causing fewer detections in regions
where the FWHM is different from the value used in the convolution and a
variable background creating fewer detections over small background and
excess detections over high background.  Thus the method used by DoPHOT and
HSTphot is more robust, providing a more uniform signal-to-noise threshold
for the star detections.

As noted in section 2.3, \textit{hstphot} allows the user to either accept
the \textit{getsky} sky image or determine a sky value adjustment before
each photometry measurement.  If the $\delta$sky option is used, the sky
modification is calculated immediately beyond the photometry radius, at a
distance of roughly 5.5 pixels from the star in PC images and 4 pixels in 
FC images.  This value, again calculated using a robust median routine, is
subtracted from the residuals during the photometry solution.  Because a
sky level determined so close to the star will invariably measure some of
the starlight as well, the PSFs are also adjusted by subtracting the
robust mean in the same region from all PSF values during the photometry
solution.

The quality-of-fit parameter that is maximized in the search for the star's
center is based on the detected signal, formal error, and $\chi$ determined
at each point.  These values are defined as follows, calculated over a
circular aperture with an effective radius of 3 pixels in the PC and 2
pixels in the WFCs.  (These radii contain $\sim$80\% of the total starlight,
using the encircled energy measurements of Holtzman et al. 1995a.)
\begin{equation}
\hbox{signal}=(\sum_{x,y} R_{x,y} \times PSF_{x,y}/\sigma_{x,y}^2 \times wt_{x,y})/(\sum_{x,y} PSF_{x,y}^2/\sigma_{x,y}^2 \times wt_{x,y}),
\end{equation}
\begin{equation}
\hbox{error}=(\sqrt{\sum_{x,y} PSF_{x,y}^2/\sigma_{x,y}^2 \times wt_{x,y}^2})/(\sum_{x,y} PSF_{x,y}^2/\sigma_{x,y}^2 \times wt_{x,y}),\ \hbox{and}
\end{equation}
\begin{equation}
\chi^2=[\sum_{x,y} (R_{x,y} - \hbox{signal} \times PSF_{x,y} )^2 / \sigma^2\times wt_{x,y}]/(\sum_{x,y} wt_{x,y}).
\end{equation}
$R_{x,y}$ is the residual after subtraction of the sky, $PSF_{x,y}$ is the
value of the PSF at the trial chip position and subpixel centering, and
$\sigma_{x,y}$ is the expected uncertainty of the measurement at that pixel,
\begin{equation}
\sigma_{x,y}^2=R_{x,y}+sky_{x,y}+\hbox{Read Noise}^2.
\end{equation}
In order to prevent $\chi^2$ from becoming extremely large for bright stars
because of PSF errors which cause $R_{x,y} - \hbox{signal} \times PSF_{x,y}$
to grow proportionally to the star's brightness, a factor of
$c^2 \hbox{signal}^2 \times PSF_{x,y}^2$ is added to $\sigma_{x,y}^2$ for
the determination of $\chi$, with $c$ values of 0.19 in the PC and 0.25
in the WFC providing a median $\chi$ near one in the final photometry.
This addition is similar to the change from Equation 1 to Equation 2 in
the cosmic ray rejection algorithm.  Finally, the $wt_{x,y}$ is a weighting
factor equal to
\begin{equation}
wt_{x,y}=R_{eff}+0.5-\sqrt{(x-x_c)^2+(y-y_c)^2},
\end{equation}
but not allowed to exceed one or drop below zero.  $R_{eff}$ is the effective
radius (3 PC pixels or 2 WFC pixels), and $x_c$ and $y_c$ are the X and Y
positions of the trial position for the star's center.

Using the calculated signal, error, and $\chi$ values, the goodness-of-fit
parameter is defined as
\begin{equation}
\hbox{fit}=\hbox{signal}/(\hbox{error} \times \sqrt{\chi^2+0.1}).
\end{equation}
This goodness of fit is essentially the signal-to-noise ratio weighted by
$1/\chi$, with the factor of 0.1 added to ensure that the equation does
not become infinite should a lucky star be perfectly-fit.  The use of this
goodness-of-fit parameter rather than a standard $\chi$ minimization is to
increase the ``capture radius,'' the maximum error in the centroid position
that would permit the star to be located by the photometric solution.  If
a $\chi$ minimization is used, an initial position estimate that falls
part way down the side of the star is as likely to slide down the profile
and locate the first available peak as it is to climb up the profile to the
peak.  However, the multiplication by the signal-to-noise will ensure that
even if the initial position estimate is well off, the star will be found.

After the entire frame has been searched for stars of all brightness
levels, all detected stars are subtracted from the residual image and the
process repeated, allowing stars located in the wings of brighter stars
to be located.  The star list is then cleaned, combining any two stars
separated by less than 1.5 pixels, and the photometry solution is begun.

\subsection{Solution and Output}

The photometry solution runs nearly identically to that of DoPHOT,
using an iterative solution.  At the start of each iteration, all stars
whose neighbors' photometry was significantly changed in the previous
iteration are flagged for solution.  ``Significantly changed'' in this
case means that one of the following conditions was true.
\begin{itemize}
\item The star's determined counts was zero.
\item The star's center moved by more than 0.3 PC pixels or 0.15 WFC
pixels.
\item The photometry changed by more than 1 count and by more than 0.0005
magnitudes.
\item The star came within 1.5 pixels of a neighbor and was combined.
\end{itemize}
The photometry of the flagged stars is then calculated, from the brightest
to the faintest, with the procedure described in the previous section used
to redetermine the positions and brightnesses.  During the solution of a
given star, all other stars in the frame are subtracted out.  The
iterative solution is considered ``partially converged'' after at least
two iterations have run and no stars were eliminated in the previous
iteration.  After the first iteration in which this is the case, the PSF
residual (described in the following section) is calculated.  The solution
is considered completely converged after a user-defined maximum number of
iterations have run or after no stars are flagged following an iteration.

This solution technique, like the initial star detection process, turned
out to be very similar to that used by DoPHOT.  However, there are a few
significant differences that should be noted.  DoPHOT uses a nonlinear
$\chi^2$ minimization routine that determines the best position and
brightness through a single minimization.  HSTphot, on the other hand,
maximizes its goodness-of-fit parameter by calculating that value
directly at a range of positions.  Given that the topology of the search
space is well-behaved for a star, such a method is advantageous in that
the determination of the goodness-of-fit at a single position is
straightforward (using the equations in the previous section) and storing
the previous trial fit values in memory will make it trivial to avoid
duplicate measurements at the a position.  Additionally, the fact that all
data are similar allows HSTphot to use a search stepsize that is
sufficiently fine to allow good fits while sufficiently large to permit a
rapid convergence.  The issue at stake is one of efficiency rather than
accuracy, with HSTphot reducing data at the same or greater speed than
DoPHOT.

An additional difference is seen when comparing HSTphot and DoPHOT with
DAOPHOT.  While the first two packages share a one-star-at-a-time solution
technique, DAOPHOT will solve for all overlapping stars simultaneously.
The speed penalty here is considerable, with an order of magnitude
difference in running speed between DoPHOT and DAOPHOT reported by
Schechter et al. (1993), thus raising the question of whether or not a
simultaneous solution of neighboring stars is actually necessary.  Given
the facts that the cores of WFPC2 PSFs are extremely narrow and that an
iterative solution can continue until the faint neighbor stars have
converged properly, it would seem that an iterative one-star-at-a-time
solution will produce equally good photometry of neighbor stars as will
the more sophisticated DAOPHOT method.  This assumption is verified
through a comparison of HSTphot and DAOPHOT photometry presented by
Dolphin (1999) for the WLM globular cluster, as well as a more
recent comparison of photometry of the Cassiopeia dwarf spheroidal
galaxy presented by Dolphin et al. (1999), with HSTphot producing sharper
CMDs with more stars in both cases.  Although it is not claimed that
the one-star-at-a-time iterative method produces better photometry, these
examples demonstrate that it will not significantly hurt the photometry,
even in the case of the very crowded WLM globular cluster field.

After convergence is reached or the maximum number of iterations have run,
a final photometry iteration is attempted to improve the accuracy.  Note
that in all previous photometry stages, star positions are only determined
at the PSF library grid points (every 0.2 pixels in the PC and 0.1 in
the WFCs).  In order to improve both astrometric and photometric accuracy,
this final iteration determines the best position to the nearest 0.01
pixels, using linear interpolation to compute the PSF at any arbitrary
position.  As with the previous photometry solutions, this initially
involves stepsizes equal to the PSF library grid spacing, but the stepsizes
are allowed to contract in this step to provide the more accurate solution.

The \textit{hstphot} output contains the following information: position,
counts, sky level, instrumental magnitude, magnitude uncertainty, $\chi$,
signal-to-noise, sharpness, and object classification.  The signal-to-noise
reported here is signal/error as determined in equations 8 and 9, divided
by $\chi$ if $\chi>1$ to provide an accurate estimate of the error.  This
signal-to-noise value is also used in determining the magnitude uncertainty.
The instrumental magnitude includes the CTE and zero point corrections that
are presented in a companion paper, although the code is included
(commented out by default) for the Whitmore, Heyer, \& Casertano (1999) CTE
corrections.

The reported sharpness is a value which would be zero if the star were
perfectly-fit, $-1$ if it were completely flat, and positive if it were
sharper than the typical star.  Its definition is
\begin{equation}
\hbox{sharp}=\frac{\sum_{x,y} ( R_{x,y}/\hbox{signal} - PSF_{x,y} ) \times (PSF_{x,y}-<PSF>) / \sigma_{x,y} \times wt_{x,y}}{\sum_{x,y} (PSF_{x,y}-<PSF>)^2 / \sigma_{x,y} \times wt_{x,y}},
\end{equation}
with $<PSF>$ being the weighted average of the PSF as defined by
\begin{equation}
\hbox{sharp}=(\sum_{x,y} PSF_{x,y}^2 / \sigma_{x,y}^2 \times wt_{x,y})/\sum_{x,y} PSF_{x,y} / \sigma_{x,y}^2 \times wt_{x,y}).
\end{equation}

Finally, the object classification comes from a comparison of the $\chi$
values of fits of the object using the best stellar profile, a single pixel
without background (ie, a hot pixel or cosmic ray), and a flat profile (an
extended object).  If either the cosmic ray or extended profiles provide
the best fit, the object is flagged accordingly.  Otherwise, one of three
star classes is used: a star detected in the first finding pass, a star
detected in the second pass and thus likely a close companion of a brighter
star, or a star combined from two stars during the photometry process and
thus likely a marginally-resolved pair.

\subsection{PSF Adjustment}

As noted in the previous section, a PSF residual image is calculated for
each chip after the first iteration in which the solution is partially
converged.  This step, which will adjust the library PSFs for the
particular focus and tracking conditions of the image being solved, is
necessary because of the implications of equations 4 and 5 - photometry
of a bright star depends primarily on the total number of counts in the PSF
(which should always be accurate) while that of a faint star depends on
the shape of the PSF.  In concept, this modification of the library PSFs
is similar to the DAOPHOT calculation of a residual used to modify its
analytic PSFs.  Thus, any error in the PSF caused by the assumption
that the library PSFs are correct in every image will have little impact
on the bright stars, but the magnitudes of the faint stars will be
systematically in error.  Stetson (1992) reported systematic errors of up
to 0.25 magnitudes over a span of $\sim$6 magnitudes in WF/PC data; my
own experiments comparing PSF-fitting photometry using unadjusted library
PSFs with aperture photometry show deviations of up to 0.15 magnitudes over
a similar magnitude range in WFPC2 images, with a typical image showing
systematic devaitions of 0.05 magnitudes.

Thus it is necessary to calculate a PSF residual image for each frame.  A
set of stars meeting the following criteria is selected as the set of PSF
stars.
\begin{itemize}
\item -0.5 $\le$ sharp $\le$ 0.5
\item $\chi$ $\le$ 4
\item The star is more than the PSF radius (6 PC pixels, 4 WFC pixels)
away from the edge of the usable chip area
\item There are no saturated or masked pixels within the PSF radius of
the star.
\item No brighter star is within 9 PC pixels or 6 WFC pixels of the
star.
\item No fainter star with at least half the brightness is within the
PSF radius of the star.
\end{itemize}

The PSF residual is then calculated through an iterative process.  First,
the average residual around the PSF stars is determined through a robust
mean at each point, comparing the residuals around the individual stars
(weighted by 1/counts, of course).  Simply adding the residual to the
current PSF, however, will not conserve the number of counts in the PSF, as
the mean residual can (and usually will) have a nonzero sum.  Since the
true PSF should be proportional to the sum of the current PSF and the mean
residual, the corrected PSF is set to be
\begin{equation}
new\ PSF_{x,y}=c(PSF_{x,y} + R_{x,y}),
\end{equation}
where $R_{x,y}$ is the mean residual at the point and $c$ is a constant
chosen to conserve the PSF size,
\begin{equation}
1/c=1+(\sum{x,y} R_{x,y})/(\sum{x,y} PSF_{x,y}).
\end{equation}
After each modification of the PSF in the iterative process, the
photometry of all PSF stars and their neighbors is recalculated.  This 
process continues until convergence is reached, with convergence defined
by no PSF point being changed by more than $\sim$0.1\% of the total
starlight.  Typically, the adjustments made to the library PSFs are small,
with the central pixels adjusted by less than 3\% of the total starlight,
indicating FWHM changes of $\sim$9\%, assuming that a typical central pixel
contains 34\% of the total light.  In a typical trial image, the addition
of the PSF residual image reduced the median $\chi$ values by 6\%.

\subsection{Aperture Corrections}

The final step necessary to have properly calibrated photometry is to
determine and apply aperture corrections, the average difference between
aperture photometry with a 0.5 arcsec radius used by H95 and the
PSF-fitting magnitudes determined by \textit{hstphot}.  It should be noted
that, since the Tiny Tim PSFs are calculated with a 0.5 arcsec radius and
normalized to a total PSF value of 1.0, and since the total PSF brightness
is conserved in both the PSF calculations and in the PSF residual
calculation, the application of aperture corrections should not be necessary
in theory.  This is actually close to the truth in practice as well, with
aperture corrections in fields with a constant background and bright,
well-separated stars generally being no more than a few hundredths of a
magnitude.  This provides the luxury of the user being able to omit the
aperture corrections altogether in cases where a lack of bright, isolated
stars would produce aperture corrections uncertain by more than this amount.

The utility provided for this task is \textit{getapcor}, which identifies
up to a user-defined number of good stars in each image with more than a
user-defined minimum number of counts.  The selection criteria for these
stars are similar to those for PSF stars, and are listed below.
\begin{itemize}
\item Must have been classified as a star
\item -0.5 $\le$ sharp $\le$ 0.5
\item $\chi$ $\le$ 2.5
\item At least 0.75 arcsec from the edge of the field or any saturated or
masked pixels
\item At least 0.75 arcsec from any brighter star
\end{itemize}
All stars that meet these criteria have the number of counts falling
within the 0.5 arcsec aperture calculated, with each value decreased by
a sky value determined in the annulus from 0.5 to 0.75 arcsec from the
star.  The aperture correction for the star is simply
\begin{equation}
\Delta mag = -2.5 \log(\frac{C_{aperture}}{C_{hstphot}}),
\end{equation}
where $C_{aperture}$ and $C_{hstphot}$ are the counts within the aperture
measured by \textit{getapcor} and the counts measured by \textit{hstphot},
respectively.  Another robust mean routine is used to determine the
overall aperture correction for each chip, based on the $\Delta mag$
values for all of the aperture stars.

\subsection{Multiphot}

A recent addition to the HSTphot package is \textit{multiphot}, a program
designed to simultaneously solve multiple images in a single run.  Again,
this concept is far from new, and is to \textit{hstphot} what ALLFRAME
(Stetson 1994) is to DAOPHOT.  The primary advantages in using
\textit{multiphot} rather than \textit{hstphot} are that it is easier to
compensate for bad pixels or cosmic rays that are only in one image and
that a more accurate photometric solution is obtained by reducing the
number of free parameters per star from $3\times N_{images}$ (X, Y, and
counts in each image) to $2+N_{images}$ (global X and Y positions and
counts determined in each image).  However, there is somewhat of a
tradeoff involved - in order to prevent \textit{multiphot} from using too
much memory (\textit{hstphot} allocates about 20 Mb per image), some
simplifications had to be made, along with the use of 32-bit floating point
rather than 64-bit.  The simplified photometry, which affects the expected
noise at each pixel, increases uncertainties by roughly 0.005 magnitudes
for bright stars, while the increased roundoff error contributes another
0.002 magnitudes of uncertainty.  Given a limiting accuracy of about 0.011
magnitudes determined for \textit{multiphot}, these uncertainties will
contribute little to the overall error budget.

Overall, the \textit{multiphot} algorithms are nearly identical to those
in \textit{hstphot}.  The only significant exception is the necessary
one: when determining the goodness-of-fit of a trial position, a combined
signal, error, and $\chi$ are used instead of just the values from a single
image.  (All PSF information is separate for the exposures, of course, with
a different residual image calculated for each image being reduced.)
However, there are a few additional complexities that must be addressed by
\textit{multiphot}.

The most significant additional problem faced by \textit{multiphot} is
that the images need to be properly aligned.  To ensure accurate PSF-fitting
magnitudes, the accuracy needs to be accurate to at least 0.1 WFC pixels.
Thus the alignment issue requires a correction for geometric distortion
(which increases the offsets between two images by up to 2\% at the corners
compared with the center), the filter-dependent plate scale (which
causes a shift of $\sim$0.4 pixels between F555W and F814W positions in
the corners), and of course an overall shift.  The Holtzman et al.
(1995a) distortion equations are adopted to determine the change in the
offset as a function of position, while the overall shift in X and Y
can be easily determined by \textit{multiphot} through a comparison of
star positions on the two frames.  No rotation is solved for, nor is any
necessary in fields that I have examined with multiphot.

The filter-dependent plate scale changes are more difficult to correct.
Initially, the Trauger et al. (1995) wavelength-dependent distortion
corrections were used to correct for both this and the geometric
distortion, but this proved to be inadequate.  Thus it was necessary to
use the empirically-determined Holtzman et al. (1995a) distortion
correction (which was calculated for F555W) and to determine the
filter-dependent corrections empirically, something that was done by
examining archival images of the Omega Centauri standard field taken in all
19 HSTphot filters (except for F170W).  The F170W correction was determined
from other archival data.  The form of the corrections was adopted from
fits to the Trauger et al. (1995) correction equations, despite the known
errors, for lack of any better source of information.
\begin{equation}
X_{filt}=(X_{F555W}-400) \times c_{filt} \times [1 + 7\times10^{-7}*((X_{F555W}-X_c)^2+(Y_{F555W}-Y_c)^2)]+400,
\end{equation}
and
\begin{equation}
Y_{filt}=(Y_{F555W}-400) \times c_{filt} \times [1 + 7\times10^{-7}*((X_{F555W}-X_c)^2+(Y_{F555W}-Y_c)^2)]+400.
\end{equation}
$X_{filt}$ and $Y_{filt}$ are the X and Y positions in any filter,
$X_{F555W}$ and $Y_{F555W}$ are the positions of the same star on an
F555W image, $c_{filt}$ is the filter-dependent term and values are given
in Table \ref{tabPS}, and  $X_c$ and $Y_c$ are roughly equal to 384 on
the PCs and 394 on the WFCs.

Finally, in order to provide a meaningful combined count total, it is
necessary for \textit{multiphot} to determine apply aperture corrections
to the count rates from the individual images.  This is done with a
routine nearly identical to \textit{getapcor}, except that it is more
forgiving of bad points in the wings of the stars.  (Given a set of many
exposures, it would likely be impossible to find any good PSF stars that
were completely free from bad pixels in all images.)

The output from \textit{multiphot} is identical in format to that from
\textit{hstphot}, except that the photometry information is listed
individually for all frames processed, as well as combined photometry
provided for every filter that was included.

\section{Tests of HSTphot}

An F555W WFC2 image of IC 1613 is displayed in Figure \ref{figRes}, with
the detected stars and cosmic rays subtracted from half the image in order
to provide in initial ``sanity check'' on HSTphot.  (Note that the white
lines and dots are the masked bad columns and hot pixels, not residuals.)
In general, \textit{hstphot} appears to be treating objects as it should.
The galaxies are left alone, aside from what are either foreground stars
or clusters or HII regions in the galaxies, which indicates that the
object classification scheme in \textit{hstphot} is working properly.

The stellar residuals themselves also provide a sanity check.  The
lack of similar features in the bright residuals is evidence that the
average PSF is correct, and that the PSF residual image was correctly
determined.  The lack of monopole residuals in the stars likewise
indicates proper PSF width and star brightness determinations.  The
lack of dipole residuals (highly positive on one side and highly negative
on the other) is indicative of proper centering.  Although this is
certainly not a quantitative test, it gives evidence that, at the
very least, HSTphot isn't completely off target in its methods.  More
detailed (and quantitative) comparisons follow in the next sections.
\placefigure{figRes}

\subsection{Photometric Reliability}

In order to determine whether or not HSTphot photometry is reliable, a
comparison was made between photometry of eight combined 2400s F555W images
of a field in IC 1613.  The data were taken with four dithering positions,
essentially providing four independent sets of observations of the same
field.  (In fact, it should be noted that the dithering provides something
of a worst-case comparison, as stars centered in one image are on the
edge or corner in the others.)  The data were processed in the standard
way, with each image combined from two 1200s images for cosmic ray removal,
and run through HSTphot independently.  (Note that \textit{Multiphot} could
have been used to produce more accurate photometry, but the resulting
photometry would not have been completely independent.)

\begin{figure}
\plotone{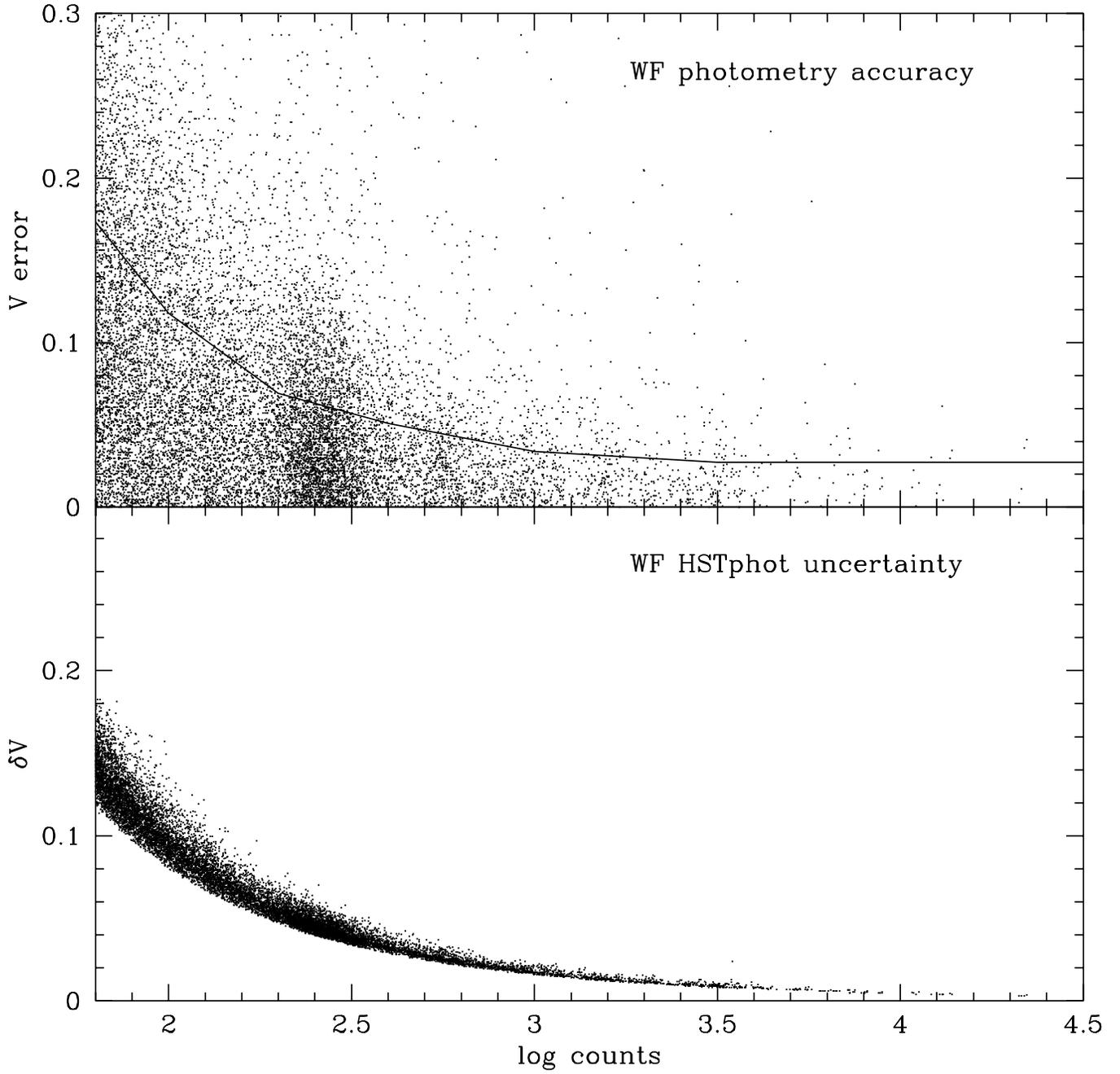}
\caption{Photometry error, based on repeated F555W observations of the IC 1613 field \label{figPhot}}
\end{figure}

The results of the comparisons between the fields are shown in Figure
\ref{figPhot}, with a 1$\sigma$ line plotted through the data.
\textit{Hstphot} uncertainties for the points are shown for comparison in
the bottom panel.  The \textit{hstphot} uncertainties match the measured
1$\sigma$ uncertainties well, except at the bright end where there appears
to be a minimum error of 0.027 magnitudes.  Part of this error is due to the
assumption that none of the stars in these data are variable, which is
incorrect.  After running a variable star analysis and retaining only the
stars that are apparently non-variable, the minimum error in the HSTphot
photometry drops to 0.02 magnitudes.  If using \textit{multiphot}, the
reduced number of free parameters in the photometric solution decreases the
error further, with a minimum of image-to-image scatter of 0.01 magnitudes
observed, consistent with Stetson's (1998) result in his CTE solution.

\subsection{Astrometric Reliability}

\begin{figure}
\plotone{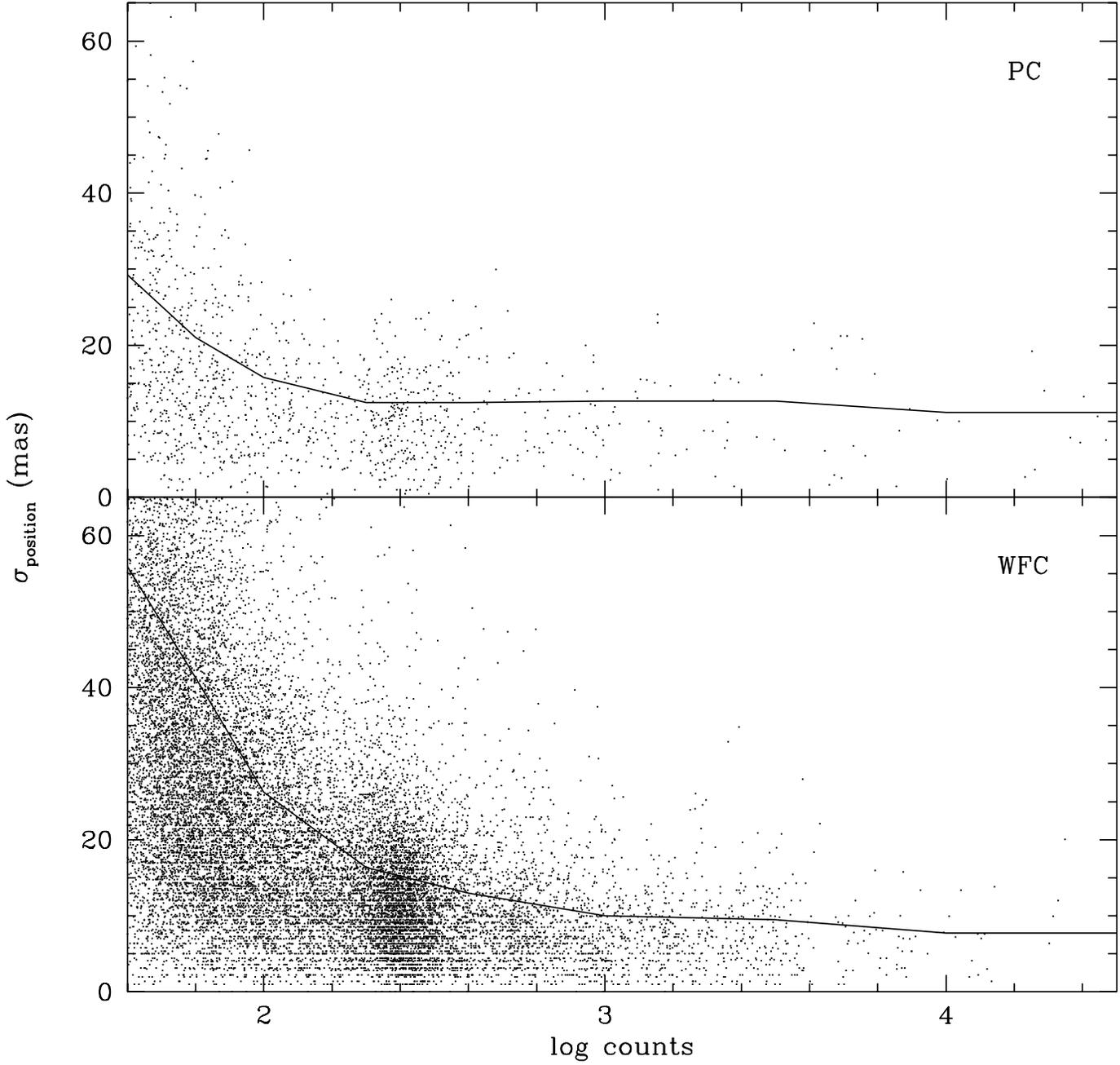}
\caption{Astrometry error, based on a global astrometry solution \label{figAstro1}}
\end{figure}

Although HSTphot was not designed to produce precision astrometry, its
fairly careful treatment of PSFs provides accurate stellar position
measurements.  The data used in the previous section can also be used to
test HSTphot's astrometric accuracy.  This was first attempted by simply
removing the Y offsets from the 34th row error, applying the Holtzman et
al. (1995a) geometric distortion equations and applying a global offset
for each star.  The results are shown in Figure \ref{figAstro1}, with a
limiting astrometric accuracy of $\sim$10 mas for the bright stars.  This
scatter is largely from the use of a global offset, rather than
independent offsets for each chip, and reflects astrometric uncertainties
in attempting a single astrometric solution for all four chips.

\begin{figure}
\plotone{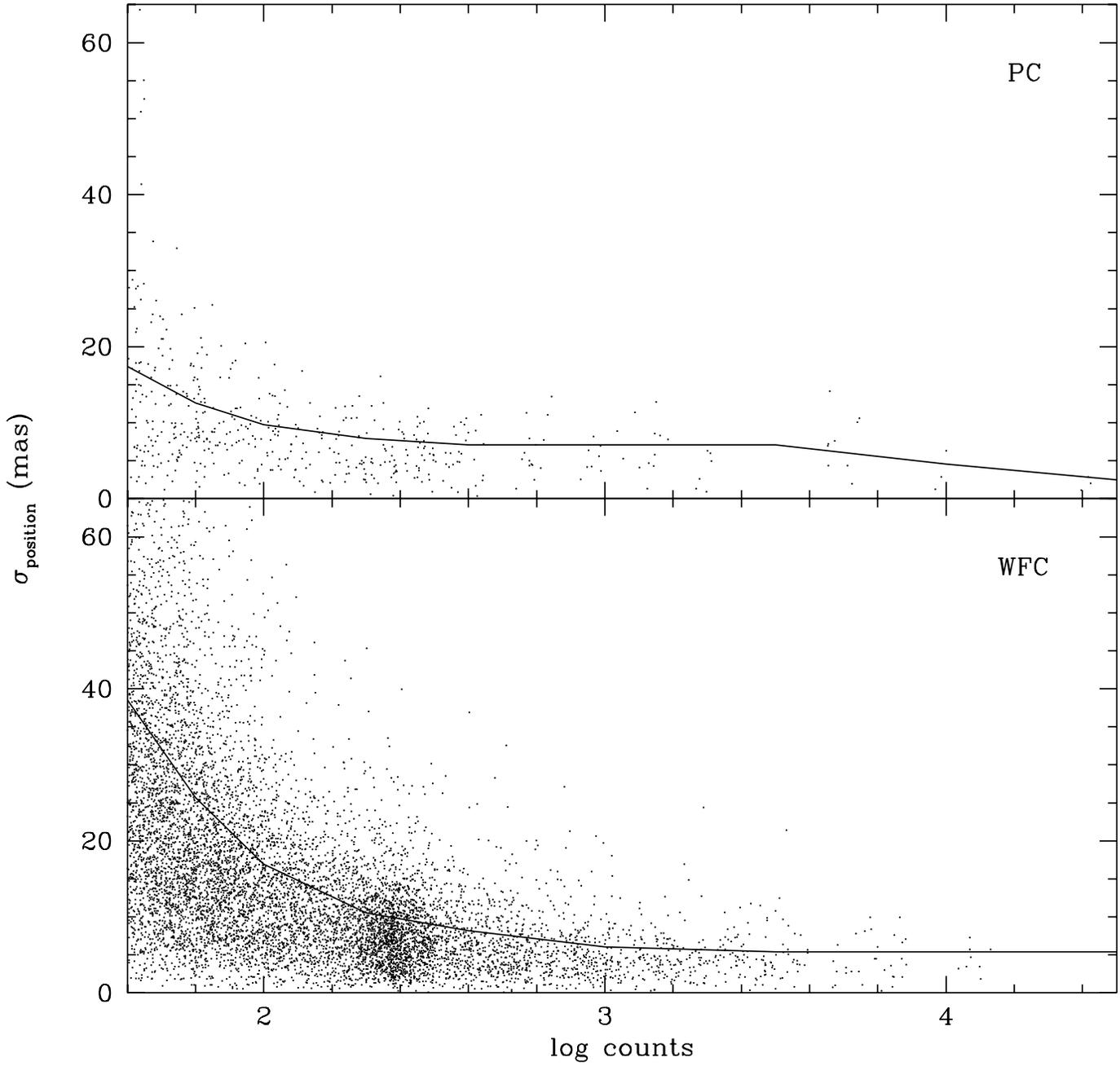}
\caption{Astrometry error, based on positions relative to a grid of bright stars in each image \label{figAstro2}}
\end{figure}

As Anderson \& King (1999) have demonstrated, high-precision astrometry
can be made by comparing star positions to a grid of nearby bright
stars.  Making a similar comparison with the HSTphot data significantly
improves the quality of the astrometry.  Results are shown in Figure
\ref{figAstro2}, with the limiting astrometric accuracy reduced to 0.05
pixels in each chip (2.5mas in the PC and 5.4mas in the WFCs).  This
value, which corresponds to 0.03 pixel accuracy in both X and Y
directions, is similar to values quoted by Anderson, King, \& Meylan
(1998).  Given that
the PSF library is only calculated with resolution of 0.2 pixels in the PC
and 0.1 pixels in the WFCs and are interpolated to intermediate values,
this accuracy is surprisingly good and could presumably be improved
further if a finer grid of PSFs were calculated.

\subsection{Object Classification}

The classification of non-stellar objects is critical to obtaining accurate
stellar photometry, as such objects would otherwise contaminate the CMD.
Three columns of the \textit{hstphot} output are useful in classification.
First is an object type, as described in Section 2.5.  It is a safe
assumption that all objects determined to be non-stellar should be omitted.
The residual shown in Figure \ref{figRes} omits extended objects, and
subtracts cosmic rays (single pixels) as such rather than as stars.

\begin{figure}
\plotone{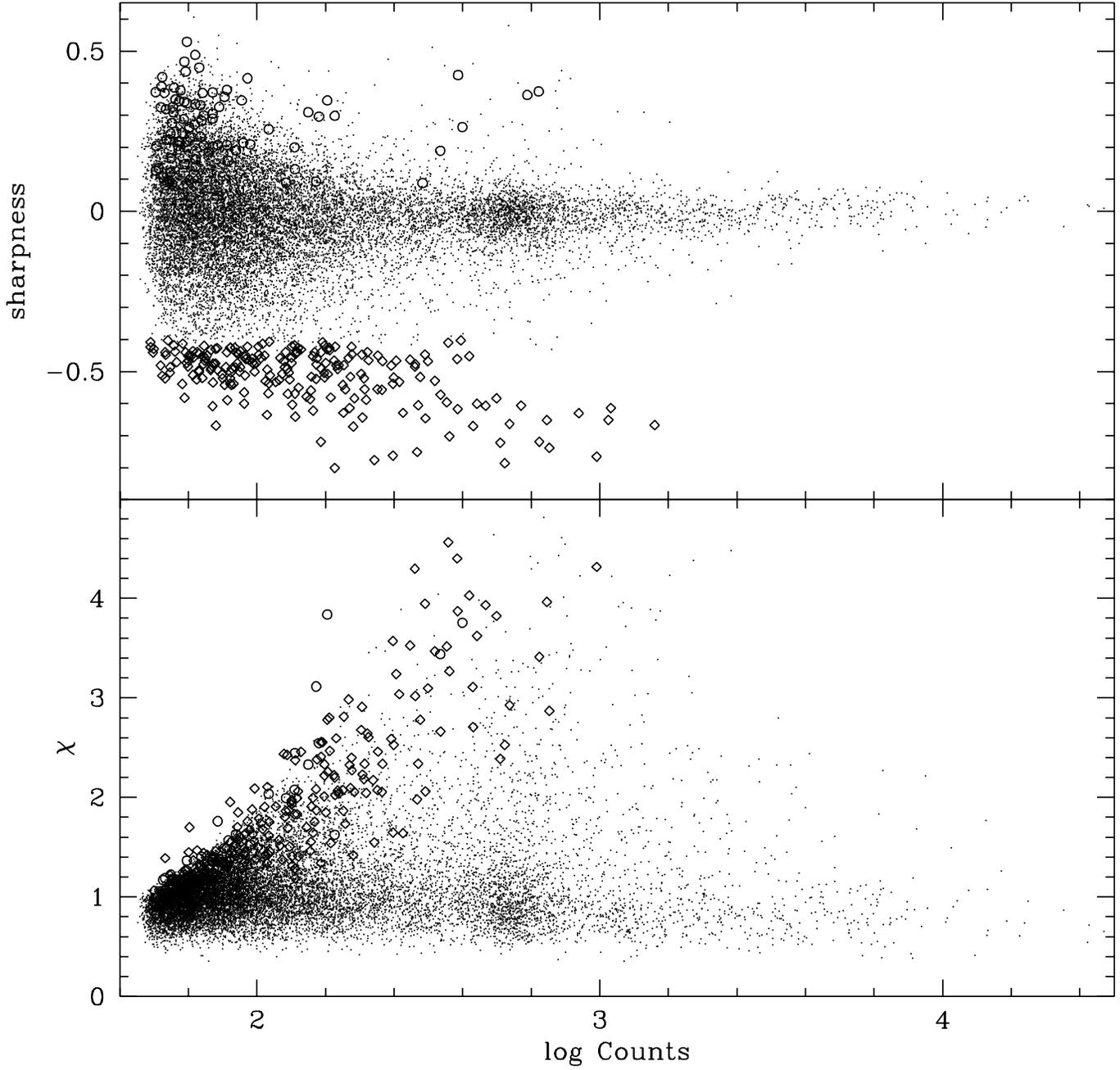}
\caption{Sharpness and $\chi$ values for stars in an F555W image.  HSTphot type 4 (cosmic rays) are plotted as circles; type 5 (extended objects) are plotted as diamonds. \label{figSharp}}
\end{figure}

The second useful column is the object's sharpness, also defined in
Section 2.5.  Figure \ref{figSharp} shows all objects in one of the IC 1613
images, with cosmic rays and extended objects (as determined by
\textit{hstphot} via object type determination) plotted with circles and
squares.  The ``stars'' well above the principal trend (at sharpness
$\ge$0.4) are multi-pixel cosmic rays, which were not identified by
\textit{hstphot} since its cosmic ray model is a single pixel.  In general,
requiring a sharpness between -0.3 and +0.3 should reject most cosmic rays
and extended objects that were not classified by \textit{hstphot}, while
retaining most of the stars.

Finally is the $\chi$ column, which simply gives the quality of the fit.
The bottom panel of Figure \ref{figSharp} shows all objects, with $\chi$
plotted against counts.  The unidentified cosmic rays seen in the
sharpness plot are also seen here, with $\chi$ values of greater than 3
or 3.5, and either limit thus useful for eliminating false detections.
The use of other poorly-fit stars depends on the quality of the data and
the application.

These plots permit the determination of selection criteria for
\textit{hstphot} output.  If one needs a complete CMD, a $\chi$ threshold
of 3 should keep most of the stars, but if a clean CMD is preferable a
$\chi$ threshold of 1.5 will eliminate most detections.  The $\chi$ limit
should be relaxed in the case of very crowded images, such as what is seen
in star clusters and in the denser regions of Local Group galaxies.  A
plot similar to that given in the bottom panel of Figure \ref{figSharp}
will quickly determine the appropriate cutoffs for any given situation.

\subsection{Comparison With DoPHOT}

Comparison photometry on one the WFC2 chip of one of the IC1613 images was
provided by Jennifer Christensen using DoPHOT, as described by Saha et al.
(1996), with aperture corrections and calibration but the no CTE or zero
point corrections applied.  Likewise the \textit{hstphot} magnitudes were
given aperture corrections but no CTE or zero point corrections.  Given
that HSTphot is somewhat of a black box with very few adjustable parameters,
my photometry using HSTphot is probably identical to what anyone else would
obtain.  The DoPHOT photometry package was run in the manner prescribed by
Saha et al. (1996).  As this method is optimized for Cepheid studies and
thus uses a background region very close to the star, the $\delta$sky
option was used in the HSTphot reduction as well to avoid an inherent
advantage given to HSTphot by the use of a larger sky region.

\begin{figure}
\plotone{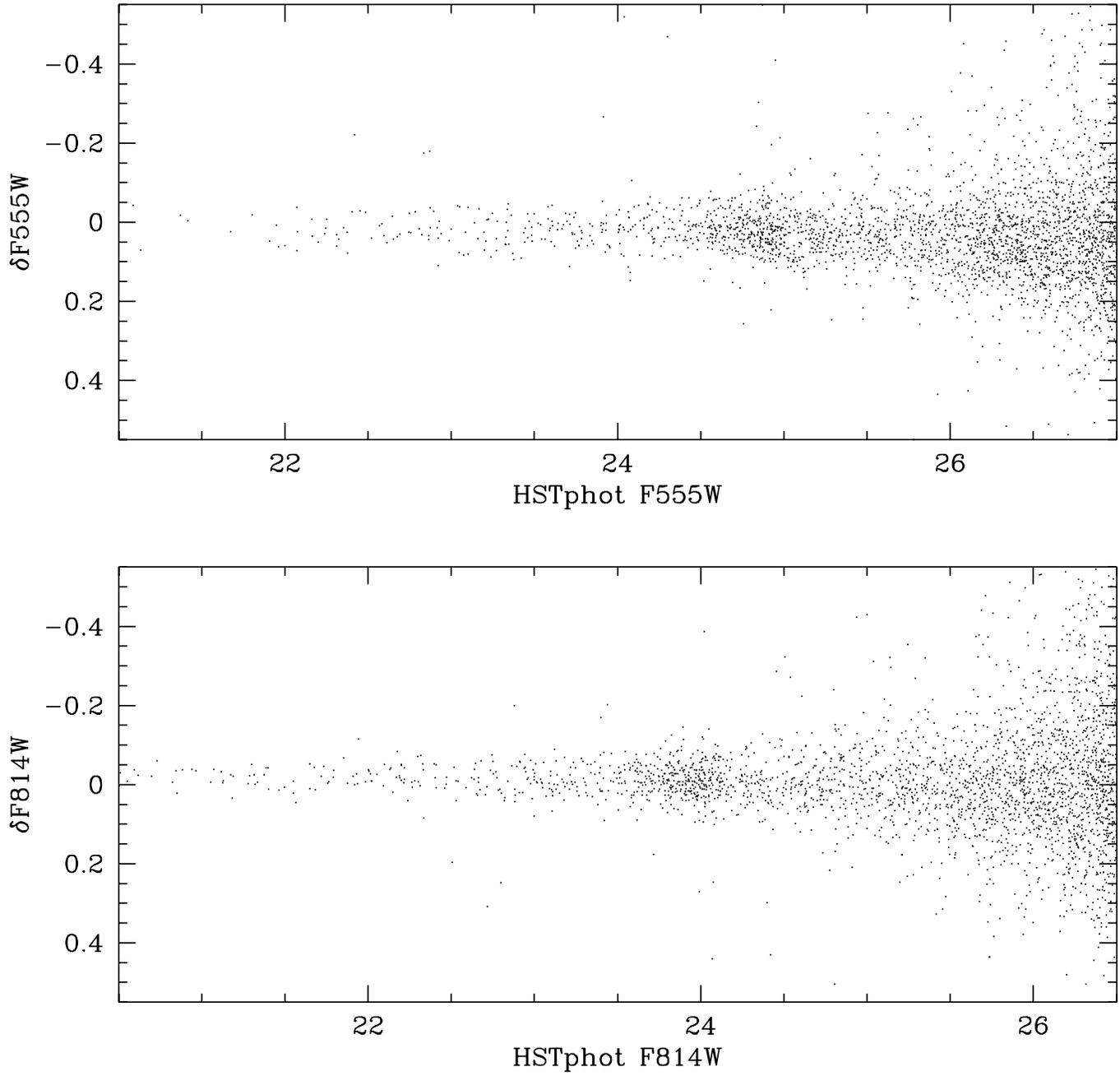}
\caption{HSTphot - DoPHOT magnitudes for the WFC2 chip of an IC 1613 image \label{figDIFF}}
\end{figure}

\begin{figure}
\plotone{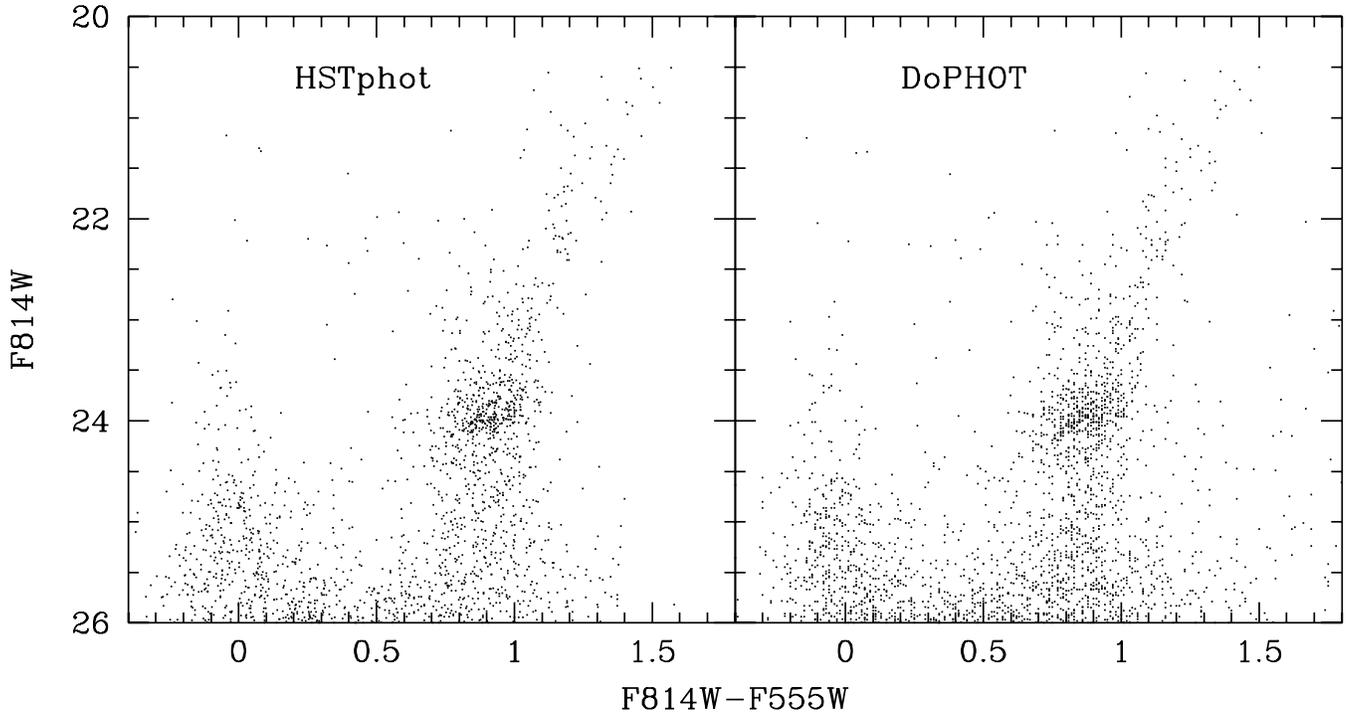}
\caption{(F555W-F814W, F814W) CMDs of the WFC2 chip of an IC 1613 image, as produced by HSTphot and DoPHOT \label{figCMDS}}
\end{figure}

The differences between the HSTphot and DoPHOT F555W and F814W magnitudes
are shown in Figure \ref{figDIFF}, and show no significant differences.
Overall, the median difference is +0.001 magnitudes, with a slight
(0.01 magnitude) difference in the individual filters that is comparable
to the uncertainty in the HSTphot aperture corrections and thus not
significant.  The respective CMDs are also shown in Figure \ref{figCMDS},
with the HSTphot CMD a few hundredths of a magnitude narrower in most
places and containing significantly fewer bad points (such as the ``stars''
to the red of the red giant branch).  Table \ref{tabCOMP} shows the average
and scatter in F555W-F814W values for two regions of the comparison CMDs -
the upper RGB ($\sim20$ stars) and the red clump ($\sim$400 stars) - with
HSTphot returning the sharper feature in both cases.  Unfortunately, the
low signal-to-noise does not permit a meaningful comparison of the main
sequence and lower RGB regions, with scatter photon noise alone being
$>$0.1 magnitudes and will thus dwarfing any minor differences caused by
the photometry programs.

\section{Discussion}

A new photometry package, HSTphot, has been developed specifically for the
challenges of reducing WFPC2 data.  Although HSTphot borrows many of its
concepts from previous reduction software, its focus on addressing one
specific problem provides for the development of a package that reduces
WFPC2 data very efficiently and accurately.  A second program,
\textit{multiphot}, is also briefly described.  This program applies
modified HSTphot routines to multiple images in order to run photometry
simultaneously.  This program will likely be the most useful in the end,
although it is newer and thus has not seen as much use (and received as
much testing) as \textit{hstphot}.  As noted in the introduction, a manual
is provided with the HSTphot package, and contains instructions for
installation of the package, preparing the data, and running
\textit{hstphot} and \textit{multiphot}.

Tests of HSTphot produced encouraging results about its photometric and
astrometric reliability.  A test of the residuals showed that
\textit{hstphot} fits PSFs, centers stars, and measures brightnesses
correctly.  It appears to have a limiting external (image to image)
photometric accuracy of $\sim$0.02 magnitudes, and an astrometric accuracy
of $\sim$0.05 pixels in position (0.03 pixels in X and Y).  A comparison
with DoPHOT reduction of a field showed no systematic differences,
although HSTphot returned a cleaner CMD with less scatter in the major
features and fewer obviously bad points.

One significant advantage in using a package designed specifically for
WFPC2 is that HSTphot runs with almost no user interaction required.  This
is possible because the number of chips, vignetted regions, rough PSF
sizes, etc. are the same in every image.  This specialization has two
benefits.  First is that the HSTphot routines are optimized for WFPC2
data, and thus run quickly.  Second is that because many of the parameters
are predetermined, HSTphot requires minimal user input and can be run
easily in batch mode.  For example, for the CTE study that I made using
HSTphot, the entire 1000+ images of Omega Centauri and NGC 2419 were
reduced at a rate of about 7.5 minutes per image, including sky calculation,
aperture corrections, and alignment.  This speed is comparable to the rate
at which the STScI archive provided the data.  HSTphot is also being used
by the HST Snapshot Survey of Nearby Dwarf Galaxy Candidates
(Seitzer et al. 1999) to provide uniform photometry of the sample of
galaxies, as well as by other users.

\acknowledgments

I would like to thank Jennifer Christensen for providing the
DoPHOT comparison photometry and Ron Gilliland for providing the data used
for the 34th row correction.  I would also like to thank Dan Zucker and
Ted Wyder for help with debugging HSTphot and facilitating the port to linux.
This work was supported by NASA through grants GO-02227.06-A and GO-07496
from Space Telescope Science Institute.

\clearpage
\begin{figure}
\caption{F555W WFC2 image of IC 1613, with stars subtracted from the left half \label{figRes}}
\end{figure}

\clearpage

\clearpage

\clearpage

\clearpage

\clearpage

\clearpage

\clearpage
\begin{deluxetable}{lrr}
\tablecaption{Filter-Dependent Plate Scale Corrections \label{tabPS}}
\tablehead{
\colhead{Filter} &
\colhead{Correction\tablenotemark{1}}}
\startdata
F170W  & 45.1 \nl
F300W  &  7.2 \nl
F336W  &  4.2 \nl
F380W  &  4.0 \nl
F410M  &  2.1 \nl
F439W  &  1.6 \nl
F450W  &  1.4 \nl
F467M  &  0.9 \nl
F547M  & -0.7 \nl
F555W  &  0.0 \nl
F569W  & -0.8 \nl
F606W  & -2.3 \nl
F622W  & -2.6 \nl
F675W  & -2.5 \nl
F702W  & -5.2 \nl
F785LP & -4.8 \nl
F791W  & -4.6 \nl
F814W  & -5.0 \nl
F850LP & -5.1 \nl
F1042M & -6.2 \nl
\enddata
\tablenotetext{1}{Values are given in units of $10^{-4}$ pixels per pixel,
and have typical uncertainties of $2\times10^{-5}$}
\end{deluxetable}

\clearpage
\begin{deluxetable}{lccrr}
\tablecaption{HSTphot-DoPHOT Photometry Comparison \label{tabCOMP}}
\tablehead{
\colhead{Region} &
\colhead{F814W limits} &
\colhead{F555W-F814W limits} &
\colhead{HSTphot F555W-F814W} &
\colhead{DoPHOT F555W-F814W}}
\startdata
Upper RGB & $21.50-22.00$ & $0.9-1.5$ & $1.197\pm0.090$ & $1.198\pm0.103$ \nl
Red Clump & $23.50-24.25$ & $0.6-1.1$ & $0.867\pm0.093$ & $0.870\pm0.095$ \nl
\enddata
\end{deluxetable}

\end{document}